# Exploring Layer Thinning of Exfoliated β-Tellurene and Room Temperature Photoluminescence with Large Exciton Binding Energy Revealed in β-TeO$_2$

Ghadeer Aljalham[1], Sarah Alsaggaf[1], Shahad Albawardi[1], Thamer Tabbakh[1,3], Frank W. DelRio[2], and Moh. R. Amer*[1,3,4]

[1]Center of Excellence for Green Nanotechnologies,
Microelectronics and Semiconductor Institute
King Abdulaziz City for Science and Technology, Riyadh, 11442, Saudi Arabia
[2]Sandia National Laboratories,
Material, Physical, and Chemical Sciences Center, Albuquerque, NM, 87123, USA
[3]Department of Electrical and Computer Engineering
University of California, Los Angeles, Los Angeles, CA, 90095, USA
[4]Department of Electrical and Computer Engineering
University of Southern California, Los Angeles, CA, 90089, USA

*Please send all correspondence to mamer@seas.ucla.edu, mamer@kacst.edu.sa

**ABSTRACT**

Due to its tunable band gap, anisotropic behavior, and superior thermoelectric properties, device applications using layered tellurene (Te) are becoming attractive. Here, we report a thinning technique for exfoliated tellurene nanosheets using thermal annealing in an oxygen environment. We characterize different thinning parameters including temperature and annealing time. Based on our measurements, we show that controlled layer thinning occurs in the narrow temperature range of 325 ºC to 350 ºC. We also show a reliable method to form β-tellurene oxide (β- TeO$_2$), which is an emerging wide band gap semiconductor with promising electronic and optoelectronic properties. This wide band gap semiconductor exhibits a broad photoluminescence (PL) spectrum with multiple peaks covering the range 1.76 eV to 2.08 eV. This PL emission coupled with Raman spectra are strong evidence of the formation of 2D β-TeO$_2$. We discuss the results obtained and the mechanisms of Te thinning and β-TeO$_2$ formation at different temperature regimes. We also discuss the optical band gap of β-TeO$_2$ and show the existence of pronounced excitonic effects evident by the large exciton binding energy in this 2D β-TeO$_2$ system that reach 1.54 eV to 1.62 eV for bulk to monolayer, respectively. Our work can be utilized to have better control over Te nanosheet thickness. It also sheds light on the formation of well-controlled β-TeO$_2$ layered semiconductor for electronic and optoelectronic applications.

**INTRODUCTION**

Emerging two-dimensional (2D) semiconductors are becoming increasingly interesting in the last few years. Efforts to identify different 2D materials for device applications can hold a great potential in solving challenges faced in semiconductor industries, ranging from efficient electronics to high performance interconnects. Transition metal chalcogenides (TMCs) have been widely studied, especially $MoS_2$, $WSe_2$, $WS_2$, $MoSe_2$ [1-6]. While TMCs exhibit a large class of materials, there are single element 2D materials that also exhibit extraordinary properties such as phosphorene, silicene, and borophene [7-13]. Among these single element materials is tellurene (Te), which is an atomically thin layer of telluride that exhibits high mobility, high thermoelectric properties, broadband photodetection, and tunable bandgap with thickness covering the range from mid to near infrared [14-18].

Tellurene can be found in different phases, most commonly α-tellurene (α-Te), β-tellurene (β-Te), and γ-tellurene (γ-Te). The latter two phases (β-Te and γ-Te) exhibit tetragonal structure, while α-Te exhibits a hexagonal structure. The α and β phases are known to be semiconductors with monolayer bandgaps of 0.75 eV and 1.747 eV, respectively [19]. β-Te exhibits anisotropy in its electronic and optical properties, which are appealing features for device applications. Accordingly, our work will focus on β-Te. Yet, most experimental investigations on layered Te is hindered by Te synthesis and growth techniques, which involves complex processes that require underpinning a controlled growth procedure [15, 20]. A less-complex method to produce layered Te makes use of a liquid exfoliation technique as the method of choice [21]. Although this method yields the desired nanosheets dimensions, it requires chemical processing which could affect layered Te properties. While liquid exfoliation may be a desirable process for specific applications (i.e., doped devices), it may not be suitable for other sensitive applications. Hence, a look for a new way to control the thickness of exfoliated Te nanosheets is sought.

In this work, we investigate controlled layer thinning of exfoliated β-Te nanosheets using high temperature annealing in oxygen. We demonstrate the different temperature regimes at which controlled layer thinning occurs. We also uncover for the first time the formation of highly-luminescent $β-TeO_2$ nanosheets. We finally discuss the results and the underlying mechanisms behind Te thinning and the formation of layered $β-TeO_2$.

## RESULTS

### β-Te nanosheet properties and lattice structure

Tellurene exhibits a trigonal one-dimensional (1D) chiral chain lattice with layers stacked on each other in a helical shape and connected via weak van der Waals forces [19]. Figures 1a and 1b show a cross-sectional view and top view of layered Te atoms, respectively. The primitive cell is demonstrated in these figures, which shows an equilateral triangle with atoms covalently connected. Figure 2a shows an optical image of a typically exfoliated Te nanosheet deposited on a $SiO_2$/Si substrate. Due to the exfoliation mechanism, the probability of obtaining a few-layer Te nanosheet is miniscule. Most of the deposited Te nanosheets exhibit a "bulk" behavior, where the thickness is significantly larger than 50 nm. This behavior is attributed to the difficulty in dry exfoliation of Te nanosheets. This can be demonstrated from AFM measurements and height characteristics in Figure 2b for the nanosheet in Figure 2a. Due to the nature of the 1D lattice, all deposited Te nanosheets exhibit a rectangular shape where one axis (in-plane *c*-axis) is exceedingly longer than the other axes. This behavior is typical of 1D lattice structures and has been observed in other 2D materials with 1D lattice structures [22].

Figure 2c shows a typical Raman spectrum obtained from a layered Te nanosheet. Two distinct Raman peaks are observed, the cross-plane $A^1$ mode and in-plane $E^2$ mode located at 120 cm$^{-1}$ and 140 cm$^{-1}$, respectively. These modes are schematically illustrated in Figures 2d and 2e, respectively. Due to the 1D anisotropy of Te, the intensities of the $A^1$ and $E^2$ modes are being modulated when angle-resolved Raman spectroscopy is measured, as shown in the polar plots in Figures 2f and 2g, respectively. This anisotropy effect is in agreement with previous reports and stems from the nature of the 1D chiral-chain lattice that Te nanosheets exhibit [21].

### Controlled layer thinning of β-Te nanosheets

To investigate Te layer thinning using oxygen annealing, exfoliated Te nanosheets were first deposited on a $SiO_2$/Si substrate. Samples were readily inserted into the furnace tube to eliminate any unwanted adsorbates and contaminations. Prior to oxygen annealing, a steady flux of argon was flown inside the chamber at room temperature and during temperature ramp up. Once the target temperature is reached, the argon flux is shut-off and a constant flux of oxygen was flown inside the chamber with a high flow rate (500 sccm). This high flow rate will ensure an oxygen-only environment inside the chamber. To investigate detrimental effects on Te layer thinning, annealing temperature and annealing time were varied. Figures S1a and S1b illustrate the thinning setup and the qualitative thinning procedure, respectively.

Figure 3 shows the morphology and optical characterizations before and after annealing. In this figure, the Te nanosheet was annealed at 325 °C for 10 min. Optical examination of the nanosheet before and after annealing was conducted, as shown in Figures 3a and 3d, respectively. A slight change in the nanosheet color is detected in the optical images which indicates light absorption has changed due to a thickness change. To shed more light on this effect, Raman intensity maps are carried out as shown in Figures 3b and 3e before and after thinning, respectively. Here, the effect of the Te morphology change can also be observed in the Raman intensity maps for the $A^1$ and $E^2$ modes, where a slight change in the intensity distribution on the surface of the nanosheet is noticed. This intensity change could also be attributed to a change in nanosheet absorption due to a thickness change, although both modes remain in the same Raman shift before and after thinning as shown in Figure 3g.

The nanosheet thinning (thickness change) can be confirmed with AFM measurements as illustrated in Figures 3c and 3f before and after annealing, respectively. The morphology of the nanosheet after oxygen annealing shows a 23 % thickness reduction, changing from 207 nm to 160 nm as illustrated in Figure 3h. We also see a change in the surface morphology, where the surface profile is non-homogeneous and exhibits a favorable thinning direction. Consequently, this thinning mechanism produces a non-uniform thickness profile and is believed to be attributed to the anisotropy of the 1D chain lattice that Te nanosheets exhibit. This thickness reduction and surface morphology profile have been consistently found in different annealed Te nanosheets at 325 °C as depicted in Figures S2 to S5.

**Complete Te thinning and formation of β- TeO$_2$**

Raising the oxygen annealing temperature will cause a rapid thinning behavior in exfoliated Te nanosheets. This is demonstrated in Figure 4 where an exfoliated Te nanosheet was oxygen annealed at 350 °C for 10 min. Figures 4a and 4d show optical images before and after this rapid thinning, respectively. A clear irreversible change is demonstrated in the optical images, with an associated color change from yellow (before thinning) to almost a transparent dark blue color (after thinning). This color change is attributed to a radical thickness reduction to a few nanometers. In fact, this rapid thinning is in agreement with the measured Raman intensity maps of the $A^1$ and $E^2$ modes in Figures 4b and 4e before and after thinning, respectively. Raman spectra in Figure 4h show the disappearance of both modes after the thinning process. The absence of these modes could be interpreted as few-layer Te nanosheets, where the $A^1$ and $E^2$ modes exhibit weak intensities for a few layers (less than 9 nm thick Te) that is attributed to enhanced interlayer interactions [15].

To understand the morphology of this rapid thinning phenomenon, AFM measurements were carried out as shown in Figures 4c and 4f before and after thinning, respectively. We see an intriguing surface thinning profile with a thickness reduction above 90 %. The annealing process at 350 °C yields a pronounced non-homogenous thickness caused by a non-uniform thinning mechanism, leading to a very rough surface. In Figure 4g, a zoomed-in plot demonstrates the surface roughness after thinning. This observed non-uniform thinning creates a mixture of 2D and 1D structures, with 1D structures extending along the thinned nanosheet 1D chain lattice axis. All samples thinned at 350 °C and above exhibit this surface thinning profile, as displayed in Figures S6 to S8. As in the case of the controlled Te thinning mentioned above, the thinning mechanism observed here is attributed to the high anisotropy effect that Te nanosheets exhibit.

The complete thinning of Te in this high temperature range also produces β-$TeO_2$. We see strong evidence of β-$TeO_2$ formation in this temperature range. This form of tellurene oxide has promising potential in semiconductor devices. In fact, recent investigations show that this form of oxide exhibits a wide and direct band gap, distinguishing itself from other $TeO_2$ phases (α-$TeO_2$ and γ-$TeO_2$) [23-25]. Nonetheless, realizing β-$TeO_2$ nanostructures has been a challenging task due to the complexity of synthesizing β-$TeO_2$ in a narrow temperature range compared to other $TeO_2$ phases [26, 27]. Consequently, this semiconductor oxide is widely understudied. Here, we report for the first-time optical band gap emission detected from oxidized Te nanosheets, which is a true signature of β-$TeO_2$.

Figure 5 shows schematics of the lattice structure of β-$TeO_2$. As shown in these figures, the lattice is 2D with an orthorhombic unit cell. A single monolayer has a thickness of 0.411 nm with a van der Waals gap of 0.201 nm. This 2D lattice structure is radically different than the 1D chain lattice found in layered Te. In Figures 6a and 6b, optical characteristics and SEM images of β-$TeO_2$ nanosheets obtained using rapid annealing of layered Te are shown, respectively. Raman spectrum in Figure 6c shows the formation of four peaks at 196.8 $cm^{-1}$, 231.6 $cm^{-1}$, 614 $cm^{-1}$, and 648 $cm^{-1}$. These Raman peaks are in alignment with the formation of β-$TeO_2$ and are classified as true signatures of β-$TeO_2$ [23, 28]. Figure S9 shows EDAX measurements of Te thinned nanosheets, which confirm the formation of β-$TeO_2$. Figure 6d shows the photoluminescence emission detected from a β-$TeO_2$ nanosheet. A broad PL emission is detected in the visible spectrum covering a wide range. This PL emission is proportional to the excitation laser parameters (intensity and exposure time). Using Gaussian fits, it is possible to fit this broad PL spectra to three overlapping peaks, namely 594 nm (2.08 eV), 651 nm (1.9 eV), and 702 nm (1.77 eV). Figure S10 shows the fitting profile for a measured PL spectrum. It is deduced that this PL emission is caused by β-TeO$_2$ optical bandgap emission. Yet, this emission falls below the predicted β-TeO$_2$ nanosheet band gap by ~1 eV. Such a behavior is attributed to the strong exciton binding energy associated with β-TeO$_2$ nanosheets, as discussed in detail below.

**Te nanosheet laser thinning and plasma thinning**

We have also explored additional methods to thin Te materials. In Figures S10 and S11, we show attempts to thin exfoliated Te nanosheets using laser thinning. This technique has been used to control layer-by-layer thinning of widely known 2D materials [29-31]. Here, the laser (532 nm) power was varied to find the minimum capable of thinning Te nanosheets. Figure S11 shows the laser heating effect on a Te nanosheet. Unlike other 2D materials, all Te nanosheets show structural damage after an optical power close to 5 mW and above, suggesting a higher optical heating sensitivity for Te nanosheets compared to other 2D materials. Thus, although this optical power is relatively low for thinning purposes, it can structurally damage Te nanosheets.

Reducing the laser power below the damage threshold does not yield noticeable thinning behavior. This is demonstrated in Figure S12. The laser power was then lowered one order of magnitude to 0.5 mW. We also varied the exposure time to induce more heating at this low power. The obtained results do not show any noticeable reduction in Te nanosheet thickness, suggesting that laser thinning of Te nanosheets is not viable. Based on these measurements, laser thinning of Te nanosheets is unreliable approach to thin Te nanosheets and can lead to structural damage and the formation of polymorph β- TeO$_2$.

We have additionally tried oxygen plasma thinning which is another technique implemented to control the thickness of layered 2D nanosheets [32-34]. Figure S13 shows optical images and AFM measurements before and after oxygen plasma treatment. Analogous to the laser thinning experiment discussed above, no noticeable plasma-induced thinning on Te nanosheets has been observed. In fact, plasma treated nanosheets exhibit the same exact thickness profile before and after plasma treatment. Accordingly, plasma induced Te thinning is not a viable approach to thin Te nanosheets.

**DISCUSSION**

Layered Te thinning using the oxygen annealing technique demonstrated here shows how to control Te layer thinning by changing the annealing temperature. To gain more insight into how temperature can affect thinning, Figure 7a plots the percentage of different Te nanosheet thinning with annealing temperature. We observe four distinctive regions in this plot: no thinning (region 1), controlled Te thinning (region 2), rapid Te thinning coupled with β-$TeO_2$ formation (region 3), and complete absence of Te nanosheet (region 4).

The "no thinning" region occurs at temperatures lower than 300 °C regardless of the annealing time. However, the narrow temperature range between 325 °C and 350 °C (region 2) is the optimal spot for well-controlled layer-by-layer thinning of Te. In this temperature range, most of the samples show thinning between ~10 % to 20 %. In contrast, the temperature range of 350 °C to 400 °C produces rapid Te thinning accompanied with β-$TeO_2$ formation. This narrow region can also be divided into two different regions based on the annealing time, as explained below. For temperatures above 400 °C, a complete absence of Te nanosheets occurs, which is expected as this temperature range is close to the melting point of Te (452 °C) [35]. Consequently, we deduce that region 2 is the region of interest for a controllable Te nanosheet thinning, while region 3 produces β-$TeO_2$ nanosheets with observable broad PL emission.

Although temperature has the largest effect on Te thinning, we also varied the thinning time of exfoliated Te nanosheets. Figure 7b demonstrates the effect of annealing time on Te nanosheet thinning for 325 °C and 350 °C. We can interpret two observations from these results. First, nanosheets that are annealed at 325 °C do not show any correlation with increasing annealing time. However, this correlation seems to differ for the 350 °C region, where a slight monotonic increase with annealing time yields more Te thinning. Samples that were thinned for 2 min exhibit on average lower thickness reduction rate compared to samples that were thinned for 10 min. Yet, the tradeoff of thinning layered Te at 350 °C will be the formation of β-$TeO_2$ on the surface of thinned Te regardless of the annealing time. Evidently, Te samples that were thinned for 2 min show a strong PL emission on their thinned surfaces as shown in Figures S6 and S7, suggesting surface oxidation. In contrast, Te nanosheets that were thinned at 325 °C exhibit a weak PL emission (Figure S14) which is comparable to the substrate background, suggesting the absence of surface formation of β-$TeO_2$. Therefore, we deduce that 325 °C is a suitable temperature for controlled Te thinning without oxidation, while 350 °C is a suitable temperature for the formation of 2D β- $TeO_2$.

As mentioned above, β-$TeO_2$ is a promising form of semiconductor oxide for a wide range of electronic and optoelectronic applications. This oxide has been theoretically studied by many groups [23-25, 36]. It was only recently that Zavabeti *et al.* demonstrated electrical characteristics of this material experimentally. First-principle calculations predict a band gap in the range of 2.2 eV to 2.6 eV [24, 36]. These DFT calculations ignore exciton binding energies and spin-orbit coupling mechanism, which can be significantly pronounced in van der

Waals materials due to quantum confinement and reduced dielectric screening [37-39]. While it might be tempting to deduce that the optical band gap (PL emission) demonstrated here is in agreement with the predicted band gap from DFT calculations, our argument would be flawed since DFT calculations do not take into account exciton effects which can be pronounced in 2D systems.

A recently reported DFT calculations carried out by Guo *et al.* predicts the accurate band gap of β-TeO$_2$, which falls between 3.32 eV for bulk and 3.7 eV for monolayer [24]. Indeed, this accurately predicted band gap range is in alignment with recent experimental investigations on monolayer β-TeO$_2$ measured using electron energy loss spectrum (EELS) and scanning tunneling spectroscopy (STS) measurements [23]. Accordingly, with the help of the measured optical bandgap emission here and using previously reported β-TeO$_2$ band gap, it is possible to estimate the exciton binding energy associated with this 2D system. We consider the optical band gap energies ($E_{op}$) which fall in the range of 1.76 eV to 2.08 eV based on PL measurements. This broad energy spectrum $E_{op}$ can be attributed to the variable β-TeO$_2$ thickness profile from bulk to monolayer and is caused by a non-uniform thinning mechanism (see AFM images in figures S6-S8). Considering β-TeO$_2$ band gap (3.3 eV to 3.7 eV for bulk to monolayer), we can readily estimate the exciton binding energy ($E_{EB}$) according to the relation ($E_g = E_{EB} + E_{op}$), which gives $E_{EB}$ values of 1.54 eV and 1.62 eV for bulk to monolayer, respectively. The energy band diagram is schematically illustrated in Figure 8 with excitonic effects considered. This obtained $E_{EB}$ is in good agreement with the universality trend between $E_{EB}$ and $E_g$ reported in literature [40, 41]. We also notice these $E_{EB}$ values are inversely proportional to the number of layers, analogous to previous reports on other 2D layered materials [42].

This exciton binding energy in β-TeO$_2$ clearly shows how PL emission can be largely shifted to lower energies than the electronic band gap. In fact, strong excitonic effects can dominate the optical response in 2D systems since weak dielectric screening coupled with low dimensionality leads to strong Coulomb interaction that contributes to a large exciton binding energy [43]. Here, the obtained $E_{EB}$ values for β-TeO$_2$ reflect the pronounced effect found in wide band gap 2D materials, which is aligned with previous predictions. To our knowledge, there are not a lot of wide band gap 2D semiconductors that exhibit strong excitonic effects with a direct band gap. We therefore believe that β-TeO$_2$ is a promising 2D wide band gap material that can potentially be used for future photonic and optoelectronic applications.

## CONCLUSIONS

In summary, we have investigated Te nanosheet thinning using oxygen annealing. We show that controlled Te nanosheets can be thinned at 325 °C for 10 min. However, raising this temperature to 350 °C will cause rapid thinning and the formation of a promising oxide β-TeO$_2$. We investigate this form of oxide optically and show for the first time PL spectra caused by optical band gap emission. This PL exhibit a wide spectral range, covering 1.76 eV to 2.08 eV. This PL emission fall below the predicted band gap for β-TeO$_2$ and is caused by strong exciton binding energy. We estimate this exciton binding energy and show a binding energy of 1.54 eV to 1.62 eV for few layers to monolayer β-TeO$_2$, respectively. Our findings can be promising for the future of electronic and photonic devices based on layered Te and 2D β-TeO$_2$.

## ACKNOWLEDGEMENTS

This research was financially supported by King Abdulaziz City for Science and Technology (KACST) through the Center of Excellence for Green Nanotechnologies (CEGN), award 20132944. T.A.T and M.R.A would like to acknowledge the support of KACST cleanroom core lab facilities. F.W.D. would like to acknowledge the support of the Center for Integrated Nanotechnologies, a Department of Energy office of Basic Energy Sciences user facility. This work was funded by the Laboratory Directed Research and Development program at Sandia National Laboratories, a multimission laboratory managed and operated by National Technology and Engineering Solutions of Sandia, LLC, a wholly owned subsidiary of Honeywell International, Inc., for the U.S. Department of Energy's National Nuclear Security Administration under contract DE-NA0003525. This paper describes objective technical results and analysis. Any subjective views or opinions that might be expressed in the paper do not necessarily represent the views of the U.S. Department of Energy or the United States Government.

## MATERIALS AND METHODS

**Nanosheet exfoliation, optical characterization, and morphology characterization**

Te nanosheets were obtained using the micro-mechanical exfoliation technique. Te crystals (2D- semiconductor) was exfoliated and deposited on a SiO$_2$/Si substrate. Confocal Raman and photoluminescence measurements (Renishaw) were taken using a 532 nm laser. A silicon detector and 2400 1/mm grating were used in the measurements. To ensure repeatability of the observed Raman spectra, multiple exfoliation processes were carried out and different nanosheets were deposited on separate SiO$_2$/Si substrates. For spatial Raman mapping, we used a streamlineHR function to measure spatial Raman intensity maps. Morphology measurements were carried out using a Park Systems XE7. We use non-contact mode to ensure no induced defects/changes to targeted Te nanosheets. Optical and AFM images were compared before and after measurements to make sure the nanosheet is still intact without any structural changes.

**SUPPORTING INFORMATION:**

Additional information on morphology characterization including AFM, Raman, and photoluminescence data of different Te nanosheets at different temperatures, EDAX images of thinned Te nanosheet, laser thinned Te nanosheets, oxygen plasma thinned Te naonsheets, and TeO2 photoluminescence fittings are all available in the supporting information.

**DATA AVAILABLILITY:** The data that support the findings of this study are available from the corresponding author upon reasonable request.

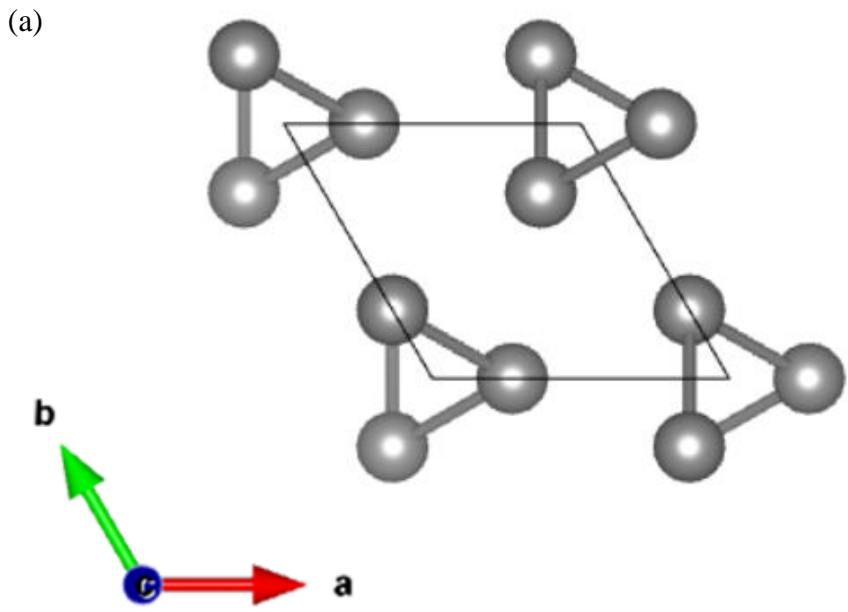

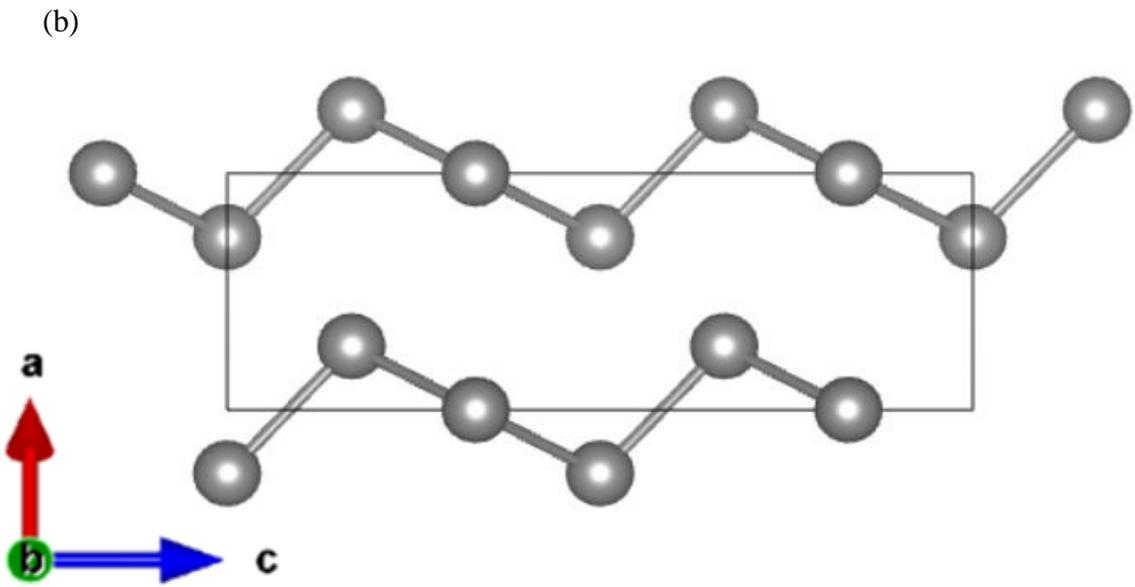

**Figure 1.** (a) Schematic diagram of the tellurene atomic structure for (a) side view (b) top view.

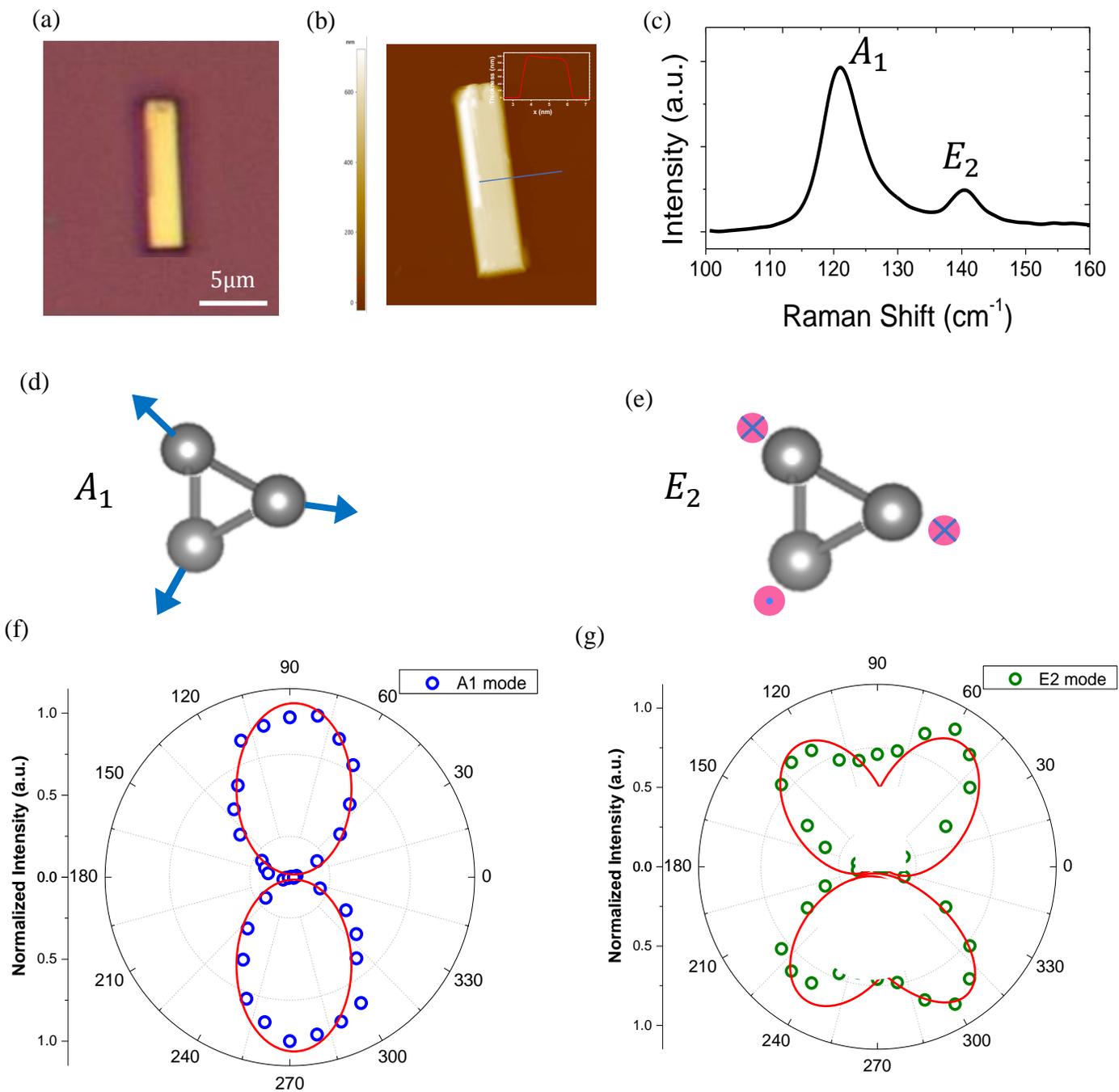

**Figure 2.** (a) Optical image (b) AFM image (c) Raman spectra of a tellurene nanosheet. The inset in (b) shows the thickness profile of the exfoliated Te. Schematics of the Raman vibrational modes for the (d) $A_1$ and (e) $E_2$ modes. Polar plots of angle resolved Raman spectroscopy for the (f) $A_1$ and (g) $E_2$ modes.

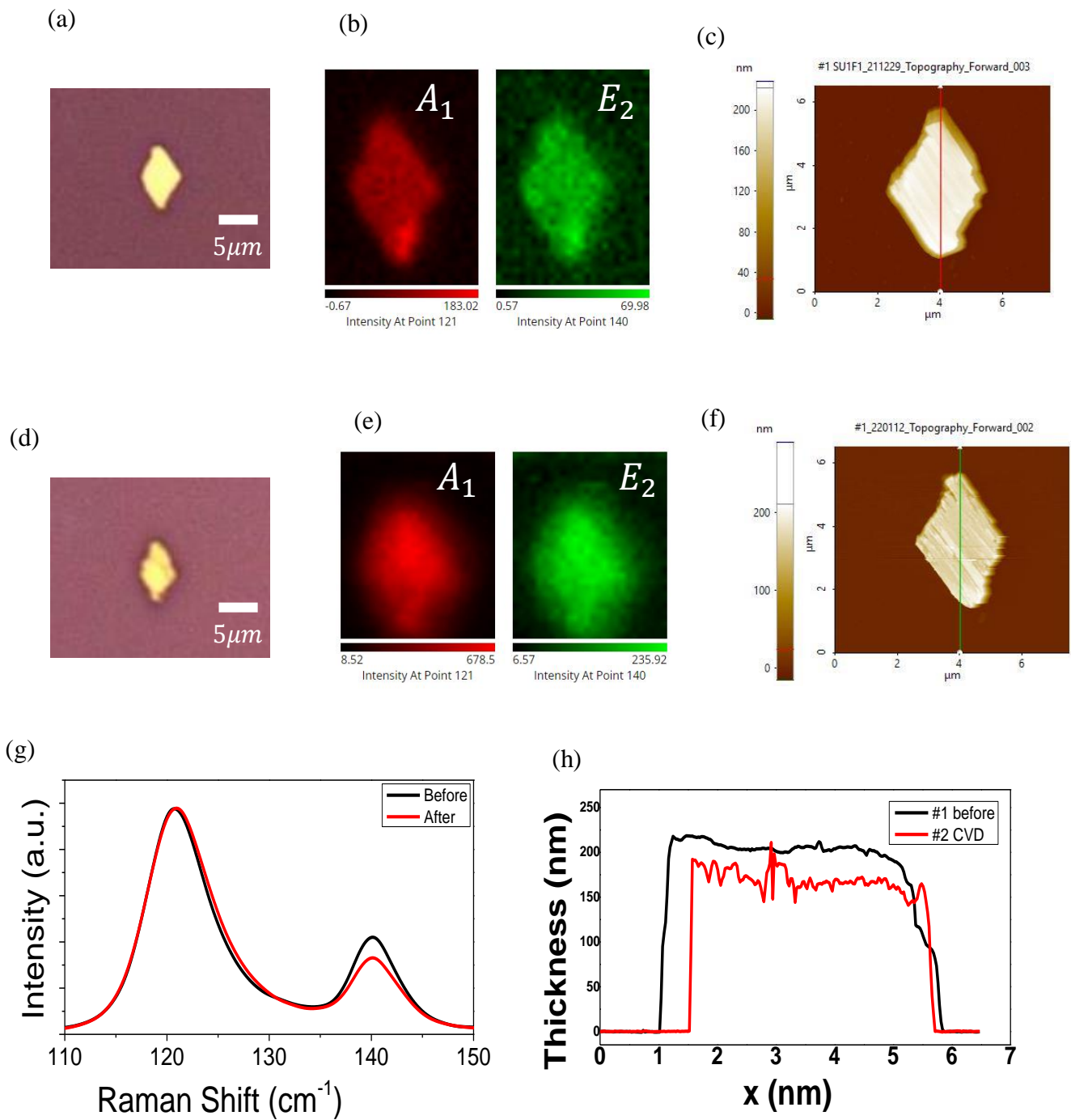

**Figure 3.** (a) Optical image (b) Raman intensity maps and (c) AFM image of a tellurene nanosheet before thinning. (d) Optical image (e) Raman intensity maps and (f) AFM image of the same nanosheet after annealing in oxygen at 325 °C for 10 min. (g) Raman spectra before and after thinning. (h) AFM height profile before and after thinning.

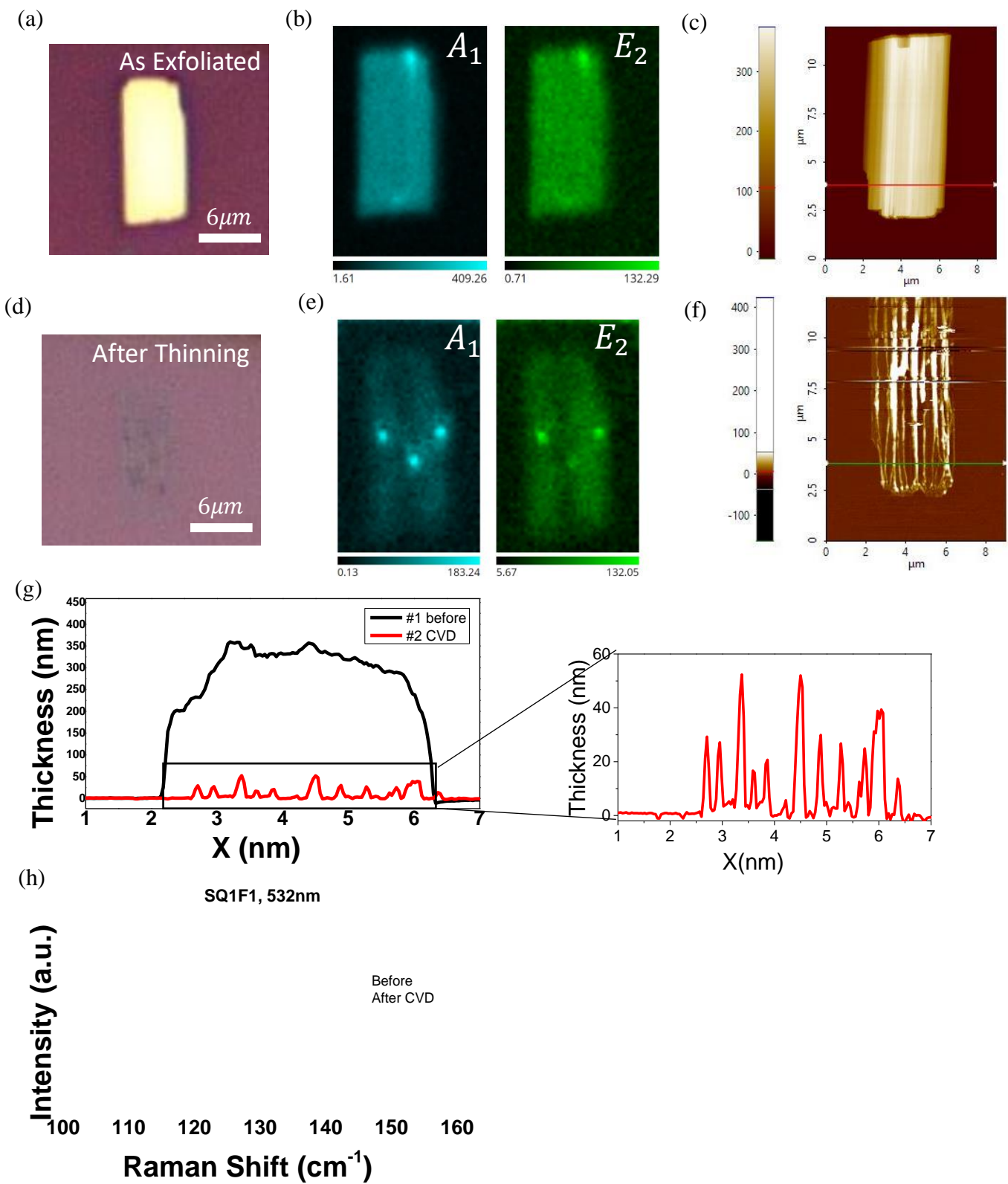

**Figure 4.** (a) Optical image (b) Raman intensity maps and (c) AFM image of a tellurene nanosheet before thinning. (d) Optical image (e) Raman intensity maps and (f) AFM image of the same nanosheet after annealing in oxygen at 350 °C for 10 min. (g) AFM height profile before and after thinning. The zoomed-in plot shows the height profile after thinning. (h) Raman spectra before and after thinning.

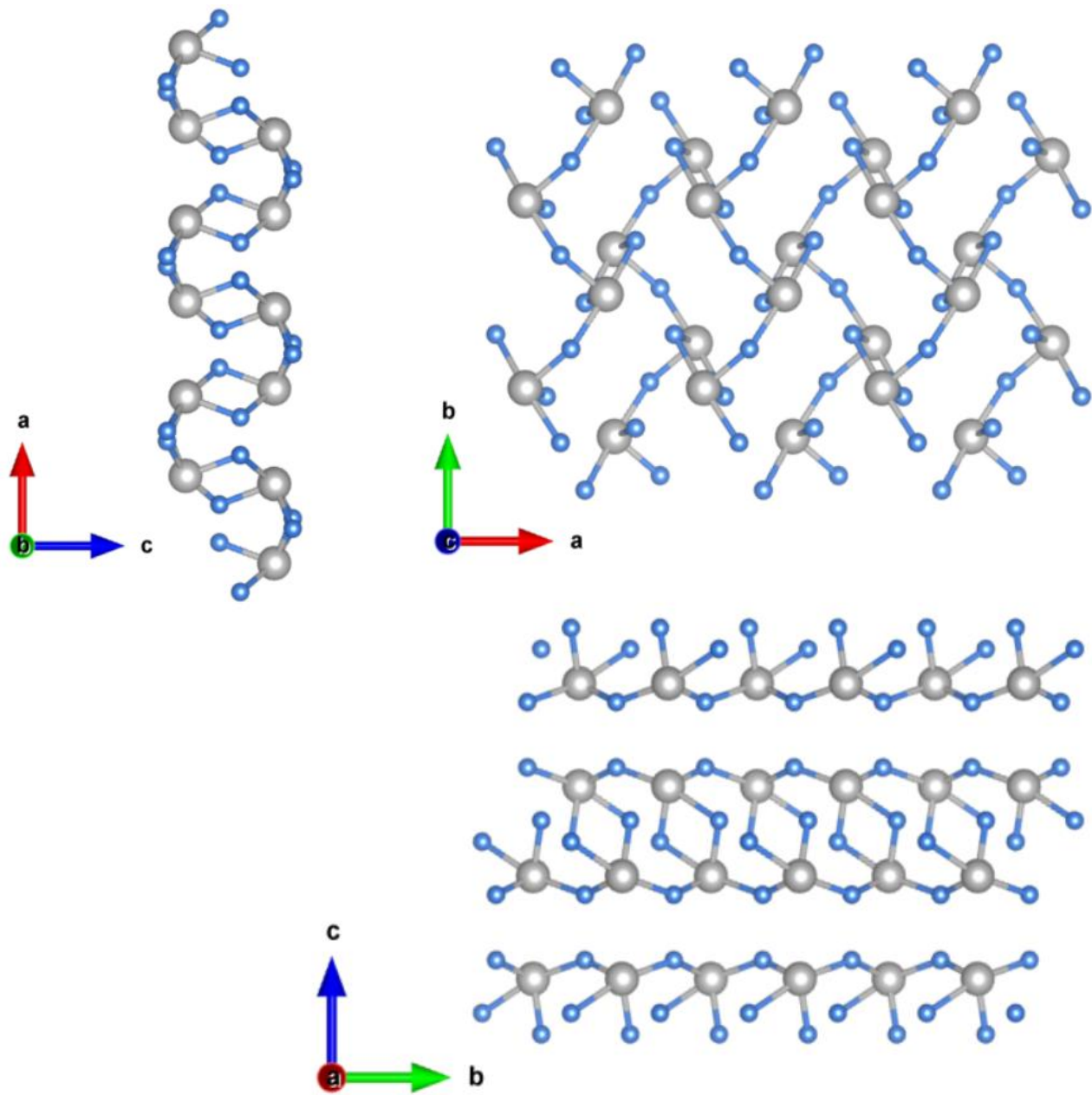

**Figure 5**. Lattice structure and the primitive cell of β-TeO$_2$ shown at different view planes. The angles of *ac* plane = *bc* plane = and *ab* plane = $\alpha = \beta = \gamma = 90°$. The unit cell parameters are a=5.54Å b=5.73Å, c=12.224Å.

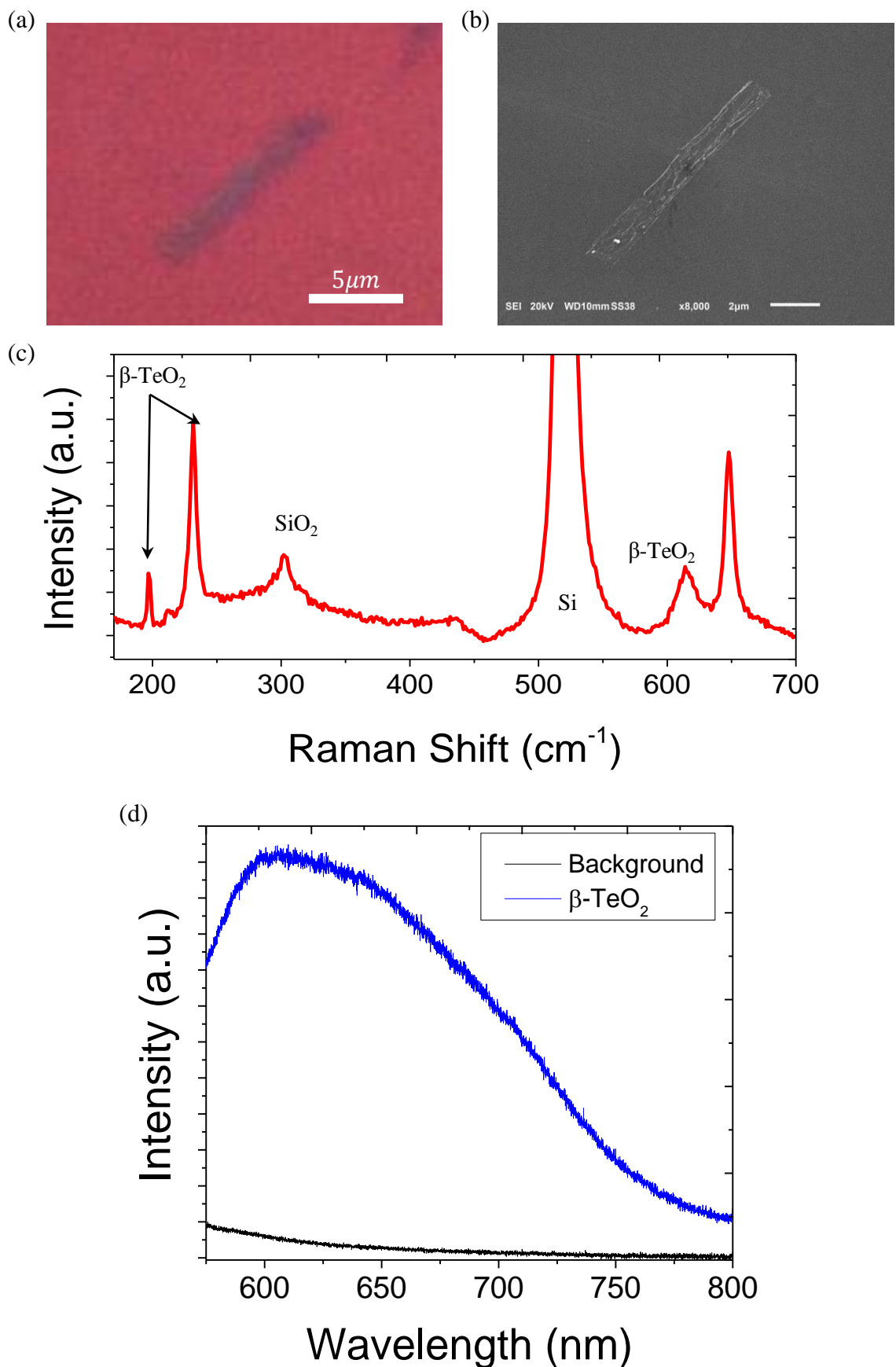

**Figure 6.** (a) Optical image and (b) SEM image of β-TeO$_2$ obtained after thinning Te. (c) Raman spectra of β-TeO$_2$ showing four signature peaks. (d) PL spectra at different laser exposure times showing the evolution of the emission.

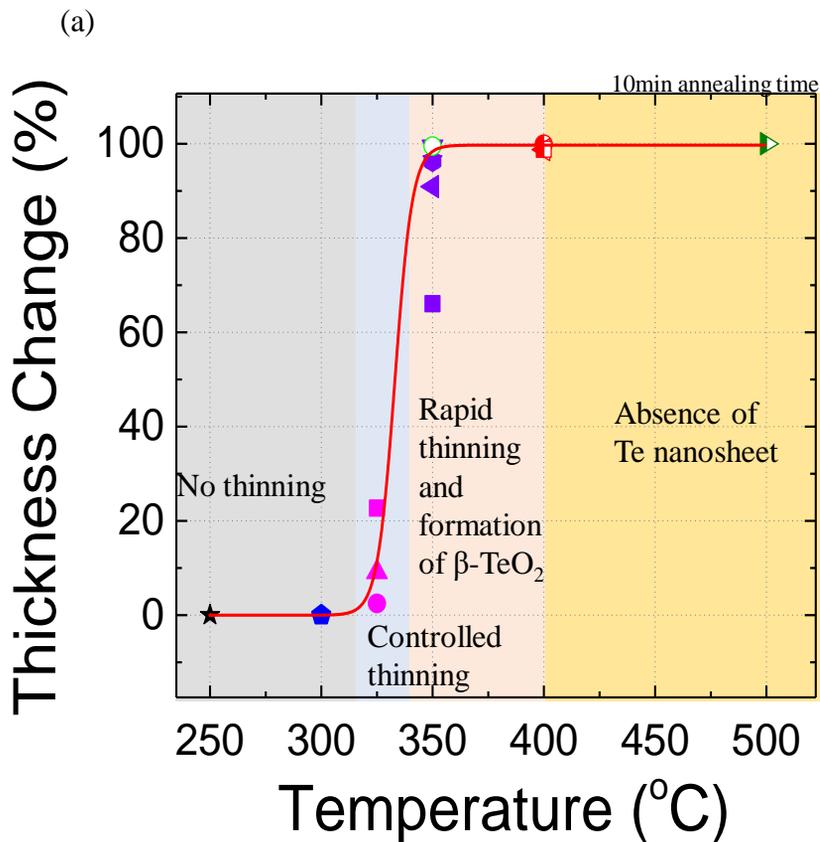

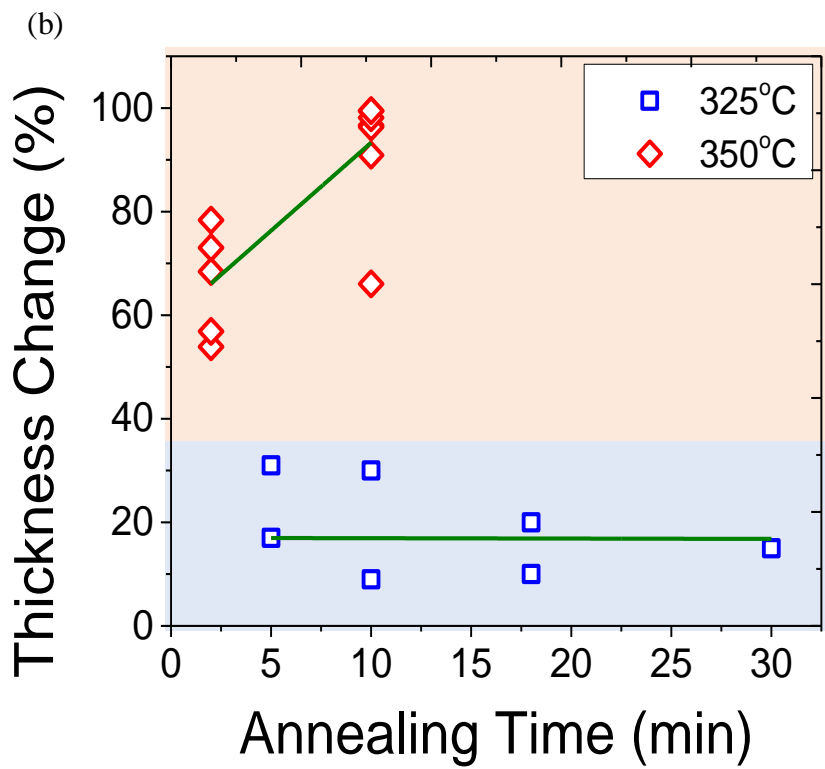

**Figure 7.** (a) Thickness change vs. annealing temperature showing four different regions, no thinning (highlighted in gray), controlled thinning (highlighted in light blue), rapid thinning and formation of β-TeO$_2$ (highlighted in light pink), and absence of Te nanosheets (highlighted in yellow). (b) Thickness change vs. annealing time for 325 °C (highlighted in light blue) and 350 °C (highlighted in light pink).

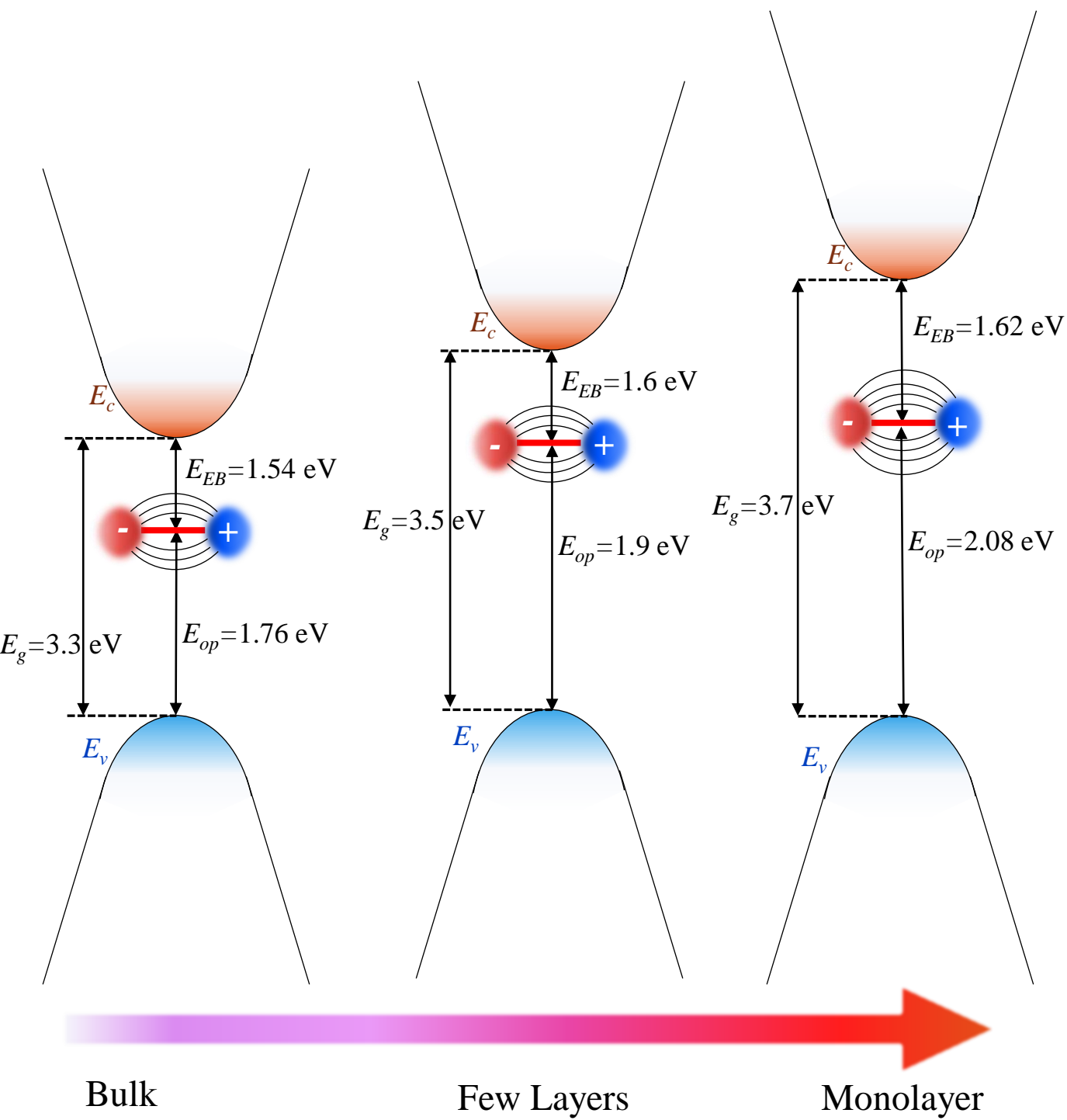

**Figure 8.** Schematic illustration showing the energy band diagram for β-TeO$_2$. The exciton binding energy ($E_{EB}$), optical band gap ($E_{op}$), and electronic band gap ($E_g$) are highlighted based on the results in the discussion.